\documentstyle[prl,preprint,eqsecnum,aps]{revtex}
\begin{document}
\tolerance=800
\draft
\newcommand{\tc}{t_{\perp}}
\newcommand{\ta}{t_{\parallel}}
\newcommand{\kpa}{k_{\parallel}}
\newcommand{\kpe}{k_{\perp}}
\newcommand{\lan}{$\mbox{La}_{2-x}\mbox{Sr}_x\mbox{CuO}_4 \;$}
\newcommand{\vecs}{{\bf S}}
\newcommand{\veci}{{\bf i}}
\newcommand{\vecj}{{\bf j}}
\newcommand{\vk}{{\bf k}}
\newcommand{\vp}{{\bf p}}
\newcommand{\vq}{{\bf q}}
\newcommand{\va}{{\bf a}}
\newcommand{\chipa}{\chi_{\parallel}}
\newcommand{\chipe}{\chi_{\perp}}
\newcommand{\mf}{{m^{\parallel}_{F}}}
\newcommand{\Mf}{{m^{\perp}_{F}}}
\newcommand{\mb}{{m^{\parallel}_{B}}}
\newcommand{\Mb}{{m^{\perp}_{B}}}
\newcommand{\Jc}{J_{\perp}}
\def\vx{{\bf x}}
\def\vy{{\bf y}}
\newcommand{\ba}{\begin{eqnarray}}
\newcommand{\ea}{\end{eqnarray}}
\newcommand{\be}{\begin{equation}}
\newcommand{\ee}{\end{equation}}
\def\ek{ \epsilon_{{\bf k}}}
\def\vek{{\bf v}\!\cdot \! {\bf k}}
\title{Coherent versus Incoherent Transport \\
 in  Layered Doped Mott Insulators}
\author{H. C. Lee$\sp{1}$ and P. B. Wiegmann$\sp{1,2,3}$
\footnote{{\tt wiegmann@control.uchicago.edu} , {\tt
hyunlee@control.uchicago.edu }}}
\address{
$\sp{1}$ James Frank Institute and  the Department of Physics \\
at the University of Chicago, Chicago, IL 60637 }
\address{
$\sp{2}$ Enrico Fermi Institute, 5640 S.Ellis Ave., Chicago, IL 60637 }
\address{
$\sp{3}$  Landau Institute for Theoretical Physics, Moscow, Russia }
\date{October 31, 1995}
\maketitle
\begin{abstract}
There exist strong experimental evidences for the dimensional cross-over
from two to three dimensions as \lan compounds are overdoped.
In this paper we  describe the dimensional
cross-over of the layered correlated metal in  the gauge theory framework.
In particular, we obtain the anomalous exponent 3/2 for the temperature
dependence of resistivity  observed in overdoped \lan.
\end{abstract}
\vskip 1.0cm
\pagebreak

\section{Introduction}
The normal state properties of high-$\mbox{T}_{c}$ compounds are anomalous.
In particular, at optimal doping in-plane resistivity $\rho_{ab}$
decreases linearly with temperature   \cite{iye}, while
out-of-plane  resistivity
$\rho_{c}$ increases with temperature. The different character of in-plane and
out-of-plane transport reflects the layered structure of the cuprates
and indicates  the hopping character of the interlayer transport.
 It is generally believed that  cuprates  evolve
into Fermi liquids as  doping increases. \\
\indent However, the  systematic studies of transport properties of overdoped
\lan  \cite{kwo1,kwo2,uchida1,uchida2} have shown a deviation from the
Fermi liquid  as well as from the optimally doped compounds.  The
temperature dependence of resistivity was found to be
$T^{\alpha}$ with an exponent close to 1.5  \cite{kwo1}. In addition
in the overdoped regime $(x \ge 0.25)$ both in-plane and out-of-plane
resistivity show similar,
although anisotropic, temperature dependence :   $\rho_c / \rho_{ab} \;$,
which is  in the order of 50 -100, is almost independent of temperature and is
 not far from the
value predicted by the  band structure calculation ($\approx 25$)
\cite{kwo1,pickett}. This
is to  be compared with the anisotropy of the order of 500-1000 at optimal
doping $x\approx 0.15 \;$ near $\, T_c \, $.
 This forces us to conclude that
 there exists a dimensional cross-over  from two dimensional {\em anomalous
(strange) metal} at  optimal doping  to  three dimensional anisotropic {\em
anomalous metal} in overdoped case.

\indent The dimensional crossover  is observed while varying the doping. In
fact, it is also a crossover in temperature.
 At sufficiently low temperature $\, T < T_{{\rm d}}(x) \,$, unless it  is cut
off by the superconducting
transition,
 any layered material is essentially three dimensional. This means, in
particular, that there is only one transport time $\tau^{-1}_{{\rm tr}}$
that determines the temperature dependence of in and out-of-plane
resistivity, so that
the ratio $\rho_c / \rho_{ab}$ does not depend on  temperature. In this case
temperature generally
increases scattering and resistivity. We  may call this type of transport
{\it coherent}.
However, at higher temperature $\, T > T_{{\rm d}}(x) \,$ all relaxation times
may  be
shorter than the interlayer hopping time, thus the  out-of-plane
conductivity is determined by  one particle tunneling.\\
 \indent Eventually, the temperature scale  $T_{{\rm d}}(x)\,$ which
determines the crossover in \\
 \lan strongly depends on the doping $\,x\sim(0.15-0.35)\,$.
The overdoped \lan  most likely lies in  the low temperature
coherent regime $ \,T < T_d(x)\, $, while  the optimally doped \lan   lies in
the high temperature two dimensional regime $ \, T > T_d(x) \, $.\\
\indent We consider the overdoped cuprates as
an intermediate metallic state which interpolates two dimensional {\em
anomalous metal} \cite{lee3}  and the conventional three dimensional metal. The
nearly
temperature independent anisotropy ratio $\rho_{ab}/\rho_{c}$ implies the
common
scattering mechanism for in-plane and out-of-plane charge transports. We
propose that
three dimensional  anisotropic gauge  theory may be a suitable model to
describe the charge transport experiments in the overdoped
cuprates as well as
 the optimally doped cuprates in an unified way.

 At $ T > T_d(x) $, when the out of-plane transport is due to one particle
tunneling, we face the question of whether an electron is a quasi-particle or
not. If
it is,  the out-of-plane conductivity is propotional to the one particle
relaxation
time $\, \tau \, $ and still decreases with temperature. In the case there
is no difference between $\, \tau_{{\rm tr}} \,$ and  $\, \tau \,$ and the
ratio $\rho_c / \rho_{ab}$
  still does not depend on temperature. If, however, electron   is not
quasi-particle due to strong interaction, (i.e. its Green function does
not possess a pole), a new time $\tau_c$ of decay to true quasiparticles
  comes into play. We call $\tau_c$  the coherence time. If the interlayer
hopping time is longer than the coherence time, the electron decays into
  components during  the hopping. We refer to this type
 of transport as {\em incoherent}. A feature of this incoherent transport is
thathe temperature increases the
 out-of-plane mobility and decreases resistivity. This situation having no
analogue in Fermi liquid  is considered in this paper\cite{note1}.\\
\indent Among  theories
proposed for the anomalous normal states of cuprates near
optimal doping, the gauge theory \cite{lee3,kotliar,wieg1}
highly emphasizes
retarded scattering by the chirality fluctuations provided by
infinitely strong on-site repulsion. In particular it gives $T$-linear
in-plane resistivity in two dimensional regime. In this paper, we extend
the gauge theory to describe out of-plane transport and the crossover
between coherent and incoherent regimes.\\
\indent We also note that  diverse models captured other mechanisms of
scattering of the peculiar out-of-plane transport of cuprates
\cite{anderson,anderson2,anderson3,kumar,legget,graf,levin,shapiro}.  \\
\indent Summarizing, we show that at  $ T \ge T_{d}(x) $ ( {\em two dimensional
regime }, optimal doping), in addition
to $\rho_{ab}\propto \; T\;,$ the gauge theory gives incoherent out-of-plane
resistivity
\be\label{1}\rho_c \propto \frac{1}{\sqrt{T}} \;, \ee

 whereas at $ T \le  T_{d}(x) $ ({\em three dimensional regime })
\be \label{2}\rho_{ab}\propto \;\rho_c\propto \; T^{3/2}\ee
The exponent $3/2$ has been observed in \lan\cite{kwo1,kwo2}.

Ironically, the gauge theory suggests  a different physical mechanism which
also gives
$\rho_{ab}\propto  T^{3/2}$  (but incoherent $\rho_c$) in two dimensional
regime. At
 high temperature the scattering by the chirality fluctuations becomes
inelastic. The inelasticity  changes the linear temperature dependence of the
resistivity to
$T^{3/2}$. Although it is unlikely that this regime  is achieved in \lan, we
discuss this mechanism in Sec.V .

\section{The Gauge Models of Normal States}
Strong on-site Coulomb repulsion forbids double
occupations  and imposes the constraint
$\sum_{\alpha}c^{\dag}_{n,\alpha}(r)c_{n,\alpha}(r)\le 1
\;$ ($r \,$ are coordinates on a layer and $n$ labels the layers).
The gauge field is a
tool to deal
with this constraint.   The constraint can be
implemented  by representing an electronic operator
$c_{n,\alpha}(r)$ by the product of a fictitious  spinon
$f_{\alpha,n}(r)$ and a holon $b^{\dag}_{n}(r)$ that keeps
track of vacant sites:
$\sum_{\alpha}f^{\dag}_{\alpha,n}(r)f_{\alpha,n}(r)+
b^{\dag}_{n}(r)b_{n}(r)=1$. One of them is a fermion, while
the other is a boson.
An accepted phenomenological model for each layer that captures a vector
character of the interaction has the form
\cite{lee1,lee3,nagaosa2,larkin,kotliar,wieg1,kalmeyer}

\begin{eqnarray}
\label{H2D}
& &H_{{\rm 2D}}= \int d^2 r \Bigl [\; \sum_{\alpha}
f^{\dag}_{\alpha}(r)\left(\,-a_0-\mu_f -\frac{1}{2 \mf} (\nabla-i \va)^2 \,
\right ) f_{\alpha}(r)  \nonumber \\
& &+b^{\dag}(r)\,\left(-a_0-\mu_b-\frac{1}{2 \mb}(\nabla-i \va)^2\right)
\, b(r) \Bigr ]
\end{eqnarray}

  A small interaction between layers can be
represented by adding an interlayer hopping term
\be
\label{H2D+}
H^{\perp}=\tc\,\sum_{n}\,\int d^2 r \,\Bigl[ \; c_{n,\alpha}(
r)c^{\dag}_{n+1,\alpha}(r)+\mbox{h.c} \; \Bigr]
\ee
We neglect the interlayer magnetic exchange
in \lan  since it is   smaller than the intralayer magnetism  by a factor of
$10^{-5}$ ( see e.g \cite{cooper} ).
We use this model (\ref{H2D}) , (\ref{H2D+}) to describe the incoherent
out-of-plane transport in the
optimally doped
 two dimensional regime, i.e. at $T > T_d(x) \;$, where the interlayer
hopping $\tc$
 is the smallest energy scale.\\
\indent At $ T < T_d(x)$ , to which we believe  the overdoped \lan belongs,
the system is assumed to be  three
dimensional  and we employ a different model which is  an anisotropic
generalization of the  two dimensional gauge theory (\ref{H2D}):

\begin{eqnarray}
\label{H3D}
 H_{{\rm 3D}}&=&\int d^3 r \,
 \, \Bigl [ \,f^{\dag}_{\alpha}(r)\,
\left(\frac{(-i\nabla-\va)_{\parallel}^2}{2\mf}+\frac{(-i\partial_z-a_z)^2}{
2\Mf}-\mu_f-a_0 \right )\,f_{\alpha}(r)  \nonumber \\
&+&b^{\dag}(r)\, \left (
\frac{(-i\nabla-\va)^2}{2\mb}+\frac{(-i\partial_z-a_z)^2}{2\Mb} -\mu_b-a_0
\right )\,b(r) \, \Bigr ]
\end{eqnarray}
In layered materials the inter-plane masses $ \, \Mb ,\, \Mf \,$ is much
larger than the in-plane masses$ \, \mb , \, \mf \, $.

\indent A few comments are  in order. The microscopic basis of the 2D model
(\ref{H2D}) is
weak, nevertheless the model has attractive universal features.  This model has
been
derived by different authors \cite{larkin,wieg1} from different physical
assumptions: In \cite{larkin}  a strong short range magnetic exchange was
essential, whereas in \cite{wieg1}
 no magnetic exchange  was assumed at all . In \cite{larkin}
$\mf$ is determined by the
magnetic exchange $J$ and $\mb$ stands for the hopping $t$. In \cite{wieg1},
both
$\mf,\mb$ are determined by the hopping. In both cases, the model (\ref{H2D})
captures  the physics
of  scattering by chirality fluctuations, namely by magnetic polarization
produced by
mobile dopants. If one is interested in how these non-local retarded processes
contribute to the normal state transport, it may be a good idea to treat
 spinon mass $\mf \,$ and  holon mass $\mb \,$
 as phenomenological parameters. \\
\indent It is even more so
for the 3D model (\ref{H3D}). The 3D model (\ref{H3D}) is suggested by
transport properties of overdoped \lan
\cite{kwo1,kwo2,uchida1,uchida2}.  We have
failed, however, in justiying this model in a quantitatively microscopic way
for the range of parameters  known for \lan. \\
\indent Another comment is that, although  two models (\ref{H2D},\ref{H2D+})
and
(\ref{H3D}) are different, they  equivalently describe the dimensional
crossover of in
plane transport.  They are essentially different, however, as far as
out-of-plane transport is concerned: while both models give the same result
for $\rho_c$
in 3D regime, i.e. at $\,T < T_d \,$, they give different $\rho_c$ at $\, T
> T_d \,$.
The reason for this is that the 3D model (\ref{H3D}) neglects  fluctuations
of the
amplitude of effective electronic hopping in-plane as well as
 between different layers
but stresses  the fluctuations of their  phases. This is a correct
approximation at low
$T$. In contrast, the 2D model (\ref{H2D}, \ref{H2D+}) neglects  the variations
of  in-plane
hopping amplitude, but takes into account the fluctuations of the out-of-plane
amplitude which  become important at high temperature.\\
\indent  At a temperature range
where  the effects of Bose condensation are irrelavant, the chirality
fluctuations are small and can be treated perturbatively. The tendency
to condense is suppressed  by the gauge interaction and strong
on-site repulsion : the holons are  hard core bosons. These
effects are beyond  the perturbation  theory and have remained  obscure. As the
result  we do not know  the low temperature bound of the perturbation
theory. Of course the  upper estimate of the bound is given by the
 mean field value of the bose condensation temperature  $T_0 \sim x/m_B$. For
cuprates $T_0$ is   too high (around 1500 K). In fact, this boson condensation
problem makes an
application of the gauge theory to cuprates questionable unless the
interactions  eliminate the condensation, thus drag  down the perturbation
theory  to much lower temperature\cite{lee3,kalmeyer}.\\
 \indent The strategy of the perturbative calculation of the transport in
the gauge
theory is well known \cite{lee1,lee3,larkin,kotliar,wieg1}.
 Let us assign electric charge to, say, the fermions.
 Then one may find  the spinon and holon currents produced by external
electromagnetic field $\, A^{{\rm ext}}_{\nu} \,$ and the gauge field:
$$j^F_\mu=\Pi^F_{\mu\nu}(a_\nu +A^{{\rm ext}}_{\nu})  , \quad
j^B_\mu=\Pi^B_{\mu\nu}a_\nu \;, $$
 where
$\Pi^{F,B}_{\mu\nu}(\vk,\omega)$ is a free fermionic (bosonic) polarization
operator. An infinite on-site repulsion, implemented by the gauge field,
renders   the spinon current to be opposite to the  holon current
$\, j^F_\mu=-j^B_\mu \,$. This allowes us to find electromagnetic current as a
response to  the external electromagnetic field
$j_\mu=j^F_\mu={\mit \Pi}^{{\rm phys}}_{\mu\nu}A^{{\rm ext}}_\nu$. The
physical conductivity
$\sigma_{\mu\nu}=\omega^{-1}{\mit \Pi}_{\mu\nu}^{{\rm phys}}(\vk=0,\omega)$  is
given by
the combination rule \cite{larkin}
 $$\Bigl({\mit \Pi}^{{\rm phys}}_{\mu\nu}(\vk,\omega)\Bigr)^{-1}=
\Bigl(\Pi_{\mu \nu}^F(\vk,\omega)
\Bigr)^{-1}+\Bigl(\Pi^B_{\mu\nu}(\vk,\omega)\Bigr)^{-1}$$\\
 At low temperature  the fermionic contribution is smaller
than the bosonic  one \cite{lee1,lee3,wieg1}. This is, roughly, due to  the
temperature dependence of the number of bosons at a given chemical
potential.
 As a result, at low temperature  the conductivity is
determined by bosonic transport relaxation time
\be
\label{sigma}
\sigma_{ab}\approx  \frac{xe^2}{\mb}\;\tau_{{\rm tr}} \, ,\quad
\sigma_{c}\approx
\frac{xe^2}{\Mb}\;\tau_{{\rm tr}}
\ee

The scalar component of gauge field is short-ranged due to Debye
screening while the  unscreened
 transverse part of vector potential produces anomalously strong
 scattering. In the next section we calculate transport time for
anisotropic 3D gauge model (\ref{H3D}) . The answer is summarized in
Eq.(\ref{2}).\\
\indent To describe  the incoherent out-of-plane transport in optimally doped
case, we
employ a different approach. In this case the interlayer tunneling is a
perturbation of the 2D model (\ref{H2D} , \ref{H2D+}).  In the lowest
order of$\,\; \tc^2 \,$  the Kubo formula gives,

\begin{equation}
\label{eq;c2}
\sigma_c^{(0)}=2\,e^2  \, \tc^2
\int_{-\infty}^{\infty}\frac{d\epsilon}{2\pi} \frac{d^2 \vp}{(2\pi)^2}
\Bigl(-\frac{\partial
n_F(\epsilon)}{\partial \epsilon} \Bigr )
\Bigl[-\frac{1}{\pi}\mbox{Im}G^R(\epsilon,\vp)\Bigr]^2 ,
\end{equation}

 where $\,G^R(x-y)\,$ is the  retarded Green function of two dimensional
electrons in a layer.\\
\indent In  Fermi liquid where one paricle Green function is
characterized  by a relaxation time
$\tau$, Eq.(\ref{eq;c2}) gives
$ \;\sigma_c^{(0)}\sim e^2\, \tc^2 \, m\, \tau \; $, and thus $ \rho_{c} $
is proportional to $\rho_{ab}$.  It appears
to be metallic and coherent even though
 $\tc$ is the lowest energy. \\
\indent The situation is very different in  the gauge theory as well as in any
other theory
 where electrons is not  quasiparticles, i.e. their Green function do not
possess a quasi-particle pole . In this case electron decay
 into "spinon" and "holon" and does not
constitute stable excitations. Spinons and holons themselves are coupled by
the gauge field and are not true  quasi-particles, either. However, at $T>T_0$
the gauge copling is weak, so that in the first approximation the electron
Green
function is simply a product of non-interacting fermion and boson Green
function.  A short range decay of the bosonic
Green function   in a layer destroys the coherence between electrons on
different layers.
 Our results for the incoherent regime is summarized in  Eq.(\ref{1}) and the
calculations are  presented in Sec.IV.\\
\indent The first step of the computation of conductivity in 3D gauge theory
(\ref{H3D} ) is to determine the propagators of the gauge fields.
Since the gauge field is  a Lagrangian multiplier, its  dynamics emerges
entirely from the polarization produced by bosons and
fermions. Perturbatively, it is given by fermionc and bosonic loops

\be
\label{P}
\Pi_{\mu\nu}(\vk,\omega)=
\Pi^F_{\mu \nu}(\vk,\omega)+\Pi^B_{\mu\nu}(\vk,\omega)
\ee
 The propagators of the gauge fields  in the transverse are the inverse of
polarization operator.
\be
\label{A}
\bigl<A_{\mu} (\vk,\omega)\; A_{\nu}(-\vk,-\omega) \bigr
>=D_{\mu\nu}(\vk,\omega)=(\Pi_{\mu\nu})^{-1}(\vk,\omega)
\ee

As in the  2D case the fermionic contribution is the
larger one, so  that only   the transversal  component of $\Pi^F_{\mu
\nu}(\vk,\omega)$ are needed. Due to the uniaxial symmetry the matrices
$D_{ij} ,\, \Pi_{ij}\,$  ($\,i,j=x,y,z \,$) can be parametrized by two
elements ($D_{\parallel},
D_{\perp}$) ( $\Pi_{\parallel}, \Pi_{\perp}$), respectively.

\ba
\label{D}
& &D_{xx}=\hat{k}_y^2
D_{\parallel}+\hat{k}_z^2 D_{\perp},\;D_{yy}=\hat{k}_x^2
D_{\parallel}+\hat{k}_z^2 D_{\perp},\;D_{zz}=(\hat{k}_x^2
+\hat{k}_y^2) D_{\perp} \nonumber \\
& & D_{xy}=-\hat{k}_x\hat{k}_yD_{\parallel},\;D_{xz}=-
\hat{k}_x\hat{k}_zD_{\perp},\;
D_{yz}=-\hat{k}_y\hat{k}_zD_{\perp}\;,
\ea
  where
\be
\label{propagators1}
D_{\parallel}(\omega,\vk)=\frac{\Pi_{\perp}k^2+\Pi_{\perp}k_z^2-
\Pi_{\parallel}k_z^2}{\Pi_{\perp}(\Pi_{\parallel}\kpa^2+\Pi_{\perp}k_z^2)}\; ,
\quad D_{\perp}(\omega,\vk)={\Pi_{\perp}}^{-1}
\ee
and  $\hat{\vk}=\vk/|\vk|$ is a unit wave vector along $\,\vk \,$, and
$\kpa^2$ is an in-plane momentum.
We assume that for typical momentum transfer $\, k_z \sim
(\Mb T)^{1/2} \, , \, \kpa \sim (\mb T)^{1/2} \, $ the following relation holds
$(\Mf d)^{-1}  k_z \ll v_F \kpa \;$, where $d$ is the inter-layer distance.
Then,
\be\Pi^R_{\parallel}(\omega,\vk)=k^2\left(\chipa- i\frac{ p_F}{ \pi
d}\;\frac{\omega}{\kpa^3}\right) \,,\quad
\Pi^R_{\perp}(\omega,\vk)=k^2 \left (\chipe -i \left(\frac{\mf}{2 d \, p_F
\,\Mf}  \right )^2\; \frac{ p_F}{ \pi d}\;\frac{\omega}{\kpa^3} \right)
\label{propagators2}
\ee
 where $ \chipa \propto 1/\mf,~\chipe \propto 1/\Mf $ are the components of
the diamagnetic susceptibilities, and  $p_F$ and $v_F$ are the
Fermi momentum and  velocity of  the two dimensional Fermi surface.
The imaginary parts  of the fermion loop are given by  the Landau damping:
\begin{eqnarray}
\mbox{Im}\Pi^R_{ij}(\omega,\vk)&=&-2\pi\omega \int \frac{d^3p}{(2\pi)^3}
v_i(\vp+\vk)v_j(\vp)\left(-\frac{\partial n_F(\xi_\vp)}{\partial
\xi_\vp}\right)\delta(\omega+\xi_{\vp}-\xi_{\vp+\vk}) \nonumber  \\
& &\approx-2\pi\mf \,\omega \int \frac{d \theta}{2\pi}\frac{d p_z}{2\pi}\;
v_{Fi}\,v_{Fj}\,\delta(\xi_{\vp_F+\vk}),
\end{eqnarray}
 where $\,v_i(\vp)=\frac{\partial \epsilon_{\vp}}{ \partial p_i },\;
\xi_{\vp}=\epsilon_{\vp}-\mu_f$,$\; {\bf v}_F={\bf v}(\vp_F) \,$ is the Fermi
velocity, $\theta $ is an angle between $\, {\bf v}_F \,$
and $\vk$, $p_F\equiv \mf v_F \;$ and  the integration over $\, p_z  \, $
 is limited by  the inverse inter-layer distance $\, \pi /d \,$.
    Employing that $(\Mf d)^{-1}  k_z \ll v_F \kpa \;$ we find that
 ${\bf v}_{F}$ is almost
perpendicular to $\vk_{\parallel}$. Under these conditions the Landau
damping is similar to the 2D case. At low $\omega < \, v_F \,\kpa \,$ we have
\begin{eqnarray}
\mbox{Im}\,\Pi^R_{yy}(\omega,\vk)&=&- \frac{ p_F}{ \pi
d}\;\frac{\omega}{\kpa}\nonumber  \\
\mbox{Im}\,\Pi^R_{zz}(\omega,\vk)&=&
-\left(\frac{\mf}{2 d \, p_F \,\Mf}  \right )^2\; \frac{ p_F}{ \pi
d}\;\frac{\omega}{\kpa}
\end{eqnarray}


\section{Anisotropic Coherent Transport}
In this section we calculate  transport time  in terms of the anisotropic
3D gauge theory (\ref{H3D}).
The calculation of  the conductivity of the system interacting via gauge forces
is peculiar. To obtain the
conductivity one must sum up the leading corrections to the vertex and
Green function of the polarization operator.  However, they are
connected by the Ward identity. This connection implements  the gauge
invariance of  interaction. Moreover in 2D  the corrections to the Green
function and to the vertex  diverge, although taken all
together , they give a finite result. Naively it looks like  there exists
a difference between transport relaxation time and one particle relaxation time
determined by the decay of one particle Green function. In fact, in our
model those  relaxation times are identical if  one takes a proper gauge
invariant definition of the one particle relaxation time,  namely as a decay of
the
 gauge invariant
Green function
$G_{{\rm inv}}(x)=\bigl < \, b(x)\exp(i\int_0^{x} a_i dx^i)b^{\dag}(0)\,
\bigr>$, being calculated on
the mass shell. Then the tail factor
$\,\exp(i\int_0^{x} a_i dx^i)\,$ takes care of the vertex corrections.
 At small  and smooth
gauge field, the gauge invariant Green function does not depend
on the path of the tail.

\indent Assuming that the gauge field is in the equilibrium,
 the relaxation time of bosons scattered by the gauge field in the second
order of the gauge field is \cite{lee3,wieg1}:
\begin{eqnarray}
\label{tau}
 \tau^{-1}_{{\rm tr}}(\vp)&\sim&\int\frac{d^{3}\vk}
{(2\pi)^{3}}\int_{0}^{\infty}\frac{d\omega}{\pi}\, {\rm Im}\langle
\Bigl({\bf v}_{\vp}\times\hat\vk\cdot {\bf B}(\omega,k)\Bigr)\;\;
\Bigl({\bf v}_{\vp}\times\hat\vk\cdot {\bf B}(\omega,k)\Bigr)\rangle
\nonumber  \\
&\times
&\Bigl(1+n_B(\omega)\Bigr)\Bigl(1+n_B(\xi_{\vp+\vk})\Bigr)\delta(\xi_{\vp}-\xi_{\vp+\vk}-\omega)|\vp|^{-2}
\end{eqnarray}
, where $\; \xi_{\vp}=\epsilon_{\vp}-\mu_B \;$.
The "magnetic field " $\, {\bf B}= \nabla \times {\bf A} \,$ is a chirality:
\be
\label{chiral}
\langle \Bigl({\bf v}_{\vp}\times\hat\vk\cdot {\bf B}(\omega,k)\Bigr)\;\;
\Bigl({\bf v}_{\vp}\times\hat\vk\cdot {\bf B}(\omega,k)\Bigr)\rangle=|\vk|^2\,
v^i_{\vp}v^j_{\vp} \,D_{ij}(\omega,\vk)
\ee
The factor $k^2$ in the in the above expression comes from the tail and
 guarantees the convergence of the scattering by soft chirality fluctuations.

The perturbation theory is valid only at temperature where  the effects of
Bose condensation are neglegible. Therefore the factor
$\,n_B(\xi_{\vp+\vk}) \,  $
can be neglected.
 For the scattering of fermion Eq.(3.1) remains the same, except  thatthe
factor
$\;1+n_B(\xi_{\vp+\vk})\;$ is replaced by
$\;1-n_F(\xi_{\vp+\vk})\;$.
 According to the  Sec.II,
the transport relaxation time of
bosons (\ref{sigma}) dominates over the
fermionic one and determines the
conductivity. \\
\indent At low temperature,  the scattering is elastic. This means that the
gauge fluctuations are damped if  the frequency $\,\omega*\sim
\,\chi\,\gamma^{-1} (\kpa)^3 \;$ ( See $\,
\Pi_{\parallel}(\vk,\omega) \,$ in (\ref{propagators2}), $\,
\gamma=p_F/(\pi \, d) \,$ ) exceeds temperature.  This
happens at $\,T< T_{{\rm in}} \equiv
(\gamma / \chi )^{2} \, (\mb)^{-3}\,$ (the opposite, inelastic case is
discussed
in sec.V ).  The out-of-plane component of the gauge field is damped at even
higher frequency $\,(\Mb/\mb) \, \omega* \,$.
 This implies that one may take into account only the {\em static} chirality
fluctuations. In static approximation Eq.(\ref{tau}) takes the form:
\be
\tau^{-1}_{{\rm tr}}(\vp)\sim T \int\frac{d^{3}\vk}
{(2\pi)^{3}} \frac{|\vk|^2}{|\vp|^2}\,  v^i_{\vp}v^j_{\vp}
\,D_{ij}(0,\vk)\,\delta(\xi_{\vp}-\xi_{\vp+\vk})
\ee

To obtain the conductivity, the momentum dependent transport time  $\,
\tau^{-1}_{{\rm tr}}(\vp) \, $ has to be averaged over the
momentum $\vp$  with the  Boltzmann distribution. The sole effect of the
averaging is to replace momentum by its thermal
value:
$p_{i}^2\sim m_B^i T,
{}~v_i^2\sim T/m_B^i \, $.
Therefore, $  \, p_{\parallel}^2 \ll
p_{\perp}^2 ~$, $v_{\parallel}^2\ \gg v_{\perp}^2~$. \\
Due to the above anisotropy  $\, k_z^2 \gg \kpa^2 \,$  holds and under this
condition Eq.(\ref{chiral}) simplifies:
\be
|\vk|^2\,
v^i_{\vp}v^j_{\vp} \,D_{ij}(0,\vk)\approx \frac{v_{\parallel}^2}{\chipa
\kpa^2+\chipe k_z^2} \,\left(k_z^2+ \frac{\chipa}{\chipe} \kpa^2 \right )
\ee
, where we kept only the term proportional to $\,v_{\parallel}^2 \,$ ,
neglecting term proportional to $\, v_{\perp}^2 \,$. The next step is the
integration over the angle between ${\bf v}_{\parallel} \,$ and ${\bf
k}_{\parallel} \,$ , which gives $\, (v_{\parallel} \kpa )^{-1} \,$. The last
integration over $ \, \kpa ,\, k_z \,$ and the thermal averaging over $\, \vp
\,$ yield  $\, \tau_{{\rm tr}}^{-1}=\tau^{-1}_{\parallel}+\tau^{-1}_{\perp}
\,$, where $\tau^{-1}_{\parallel} \,$ and $\tau^{-1}_{\perp} \,$ are given by:

\begin{eqnarray}
\label{eq:taupa}
& &\tau_{\parallel}^{-1}\sim \frac{T\,\sqrt{\Mb T}}{\chipa \mb}
\;  \\
\label{eq:taupe}
& &\tau_{\perp}^{-1}\sim\frac {T\,\sqrt{\Mb T}}{\chipe \Mb}  \nonumber \\
\end{eqnarray}
The essential difference of the above result with two dimensional one is an
extra factor $\sqrt{ T}$, which originates from the density of states.
We acknowledge that $ \, T^{3/2} \, $ dependence of the resistivity in the
context of gauge theory was  mentioned in \cite{kotliar}. \\
\indent There is a simple way to understand Eqs. (\ref{eq:taupa}) and
(\ref{eq:taupe}):
$\, \tau_{\perp}, \;\; \tau_{\parallel} \,$  contain the static  chirality
fluctuations
$\Bigl<B_{x}^2(\vk) \Bigr >\sim  \Bigl<B_{y}^2(\vk) \Bigr > \approx T/\chipe ,
\;
\Bigl<B_{z}^2(\vk)
\Bigr > \approx T/\chipa $ ,the  projected area onto the xy- plane and
yz-plane of the contour composed of the path of a boson in a unit time:
$S_{yz} \sim (\mb\Mb)^{-1/2},\;\,S_{xy}\sim  (\mb)^{-1}\,$ and the density of
states in the parallel and transverse directions: $(\mb T)^{1/2} ,\; (\Mb
T)^{1/2}$. Considering the products of three factors,  $\tau^{-1}_{\perp} \,$
and $ \tau^{-1}_{\parallel} \,$ can be obtained, respectively.

\indent As  discussed in the introduction, the anisotropic
 gauge theory (\ref{H3D})  is assumed to be valid at
temperature below the dimensional crossover temperature.
However, the 3D theory can give an upper limit  for the crossover temperature
$T^{\perp}_d$.
 Interlayer relaxation rate  $\tau^{-1}_{\perp}$ increases with the anisotropy.
 When  it reaches the interlayer hopping
 amplitude $\tc$,
the  kinetic equation and , as a consequence,
Eq.(\ref{eq:taupa},\ref{eq:taupe}) are no longer valid.
 Thus the out-of-plane conductivity reverses its temperature behaviour
(see  a footnote in introduction).
It is likely that at $\, T=T_{{\rm d}}^{\perp} \,$ the out-of-plane wavelength
 $~1/p_{\perp}\sim (\Mb T)^{-1}$ reaches the inter-layer distance $d$.
 Then, the condition $\tau^{-1}_{\perp}\sim \tc$ gives a
temperature scale of the crossover $T_{{\rm d}}^{\perp}=\tc\,\Mb\,d\,\chipe~$.
If the
 value of $\Mb$ can be identified with $\, (\tc d^2)^{-1} \,$, $T_{{\rm
d}}^{\perp}\sim \chipe/d$.\\
\indent On the contrary $\tau_{\parallel}^{-1}$ is smooth through dimensional
cross-over, and Eq.(\ref{eq:taupa}) is still valid for in-plane transport.
The only
difference
is that the integration over $p_z$  have to be cut off by the inverse
interlayer distance $1/d$. Therefore, at
 $T > (2 \Mb d^2)^{-1} $:
\begin{equation}
\tau^{-1}_{\parallel}\sim\frac{T}{\chipa \, \mb} \, \frac{1}{d}
\end{equation}
This is  the well-known $T$-linear in-plane resistivity in two dimensional
limit. Note that  in this limit $ \, \chipa \, d = \chi^{{\rm (2D)}} \, $.\\
\indent Observe that the dimensional cross-over of the   in-plane and
out-of-plane
transport stem from different mechanisms and may occur at
different temperatures.
Nevertheless, if one assumes that all phenomenological parameters of the
out-of-plane part of the model are of the same order
$\,(\Mb \,d^2)^{-1}\sim \chipe/d  \sim t_{\perp}$  the estimate of the
crossover temperature is
\be\label{T_d}
T_{{\rm d}}(x)\sim t_{\perp}
\ee
\indent  Let us discuss the experimental side of the story
\cite{kwo1,kwo2,uchida1,uchida2}.
The experiments have shown that in overdoped \lan $\rho_{ab}$ is proportional
to
$T^{\alpha}$, where $1<\alpha <2$ \cite{kwo1,kwo2,uchida1,uchida2}.
Notably in Ref.\cite{kwo1} $~\alpha$ was found to be very close to 3/2
for \lan with $x$=0.35 and agrees with  the  Eq.(\ref{2}). \\
\indent From the data of Ref.\cite{kwo1} the experimental value of
$T_d(x=0.35)$ can be estimated to be around 800 K. Measurements of the
c-axis polarized
optical spectrum over doping range $0.1 < x < 0.3~$
 \cite{uchida2}  are consistent with resistivity data.
 An  estimate of $\tc$ may be taken from  the optical conductiviy data.
The  Drude-like  fitting
gave $\tau_{\perp}^{-1}~\sim \max(\omega^{\alpha},T^{\alpha}) $ with $1<\alpha
<
2$. In \cite{uchida2} the ratio of in-plane and out-of-plane  plasma
frequencies was also  found . At doping $x=0.15$ and $x=0.3$,
$\omega_{p,\parallel}/\omega_{p,\perp} \approx 30$ and $10$, respectively.
These data enable one to get an estimate of  $\tc$ \cite{cooper,kwak}:
$$\frac{\omega_{p,\perp}}{\omega_{p,\parallel}}=\sqrt{2}\left(\frac{d}{a}\right)\frac{\tc}{E_a^{\prime}} \;, $$
where $a=3.79 \: \mbox{\AA}$ is the lattice spacing
in layer and $d=13.21\: \mbox{\AA}$ is  the inter-cell distance of \lan.
$E_a^{\prime}\,a = \hbar \, v_F \;$ and $v_F$ is the in-plane Fermi
velocity \cite{kwak}. According to
\cite{allen} , the band structure calculations yield $\, v_F=3.1 \times 10^7
\mbox{cm/s} $ at $\, x=0.15
,\,x=0.20 \,$. Combining all of the above formula and data  we obtain a
somewhat lower $\tc\sim 200\: \mbox{K}\;$
for
 $x=0.3-0.35$.
\\
Near the optimal doping \cite{cooper} $\, \tc \sim 2.4  \:\mbox{meV}=28
\:\mbox{K} \:$ at $\, x=0.16 \,$ , which is also obtained from the optical
measurements. The superconducting transition temperature at $\, x=0.16 \,$
is $\, T_c(x=0.16)=34 \:\mbox{K} \:$. Thus near optimal doping the
dimensional cross-over can be possibly screened by the superconducting
transition. \\
\indent In the overdoped case  we may rely
on the band theory.  The value of  hopping amplitudes
quoted in Ref.\cite{anderson} are $~\ta\approx 0.5 \:\mbox{eV} \sim 6,000 \:
\mbox{K}$ and $\tc \approx 0.05 \:\mbox{eV} \sim 600 \:\mbox{K} \;$  The
band theory value of $\tc \,$ is not very different from the value of the $
\, \tc \, $ obtained  above from optical data \cite{uchida2}. This can be
expected , since  an
interaction in overdoped case is not  as strong as in the case
 of optimal doping.\\
\indent Above estimates of $T_{{\rm d}}(x) \,$ indicates that there is a room
for the three dimensional regime in overdoped  \lan ($x$ $>$ 0.25).

\section{Incoherent Transport in Optimally doped \lan}
\label{sec;inco}
In this section we consider the out-of-plane transport
in  case  the interlayer hopping amplitude $\tc$ is the smallest rate:
(i) the time of hopping $\tc^{-1}$ is longer than in-plane relaxation time and
 (more importantly)
(ii) longer than the characteristic time of all kind of  magnetic
fluctuations.
This case corresponds to the optimally doped \lan.\\
\indent Under the condition (i) the hopping term (\ref{H2D+}) can be treated as
perturbation and
under the condition (ii) the approximation which allows us to write the
hopping term (\ref{H2D+}) in the form (\ref{H3D}) is no
longer valid.\\
 \indent The c-axis conductivity was calculated with 2D gauge theory  in
\cite{nagaosa2,anderson}.  In \cite{anderson}, $\rho_{c} \propto 1/T$ was
obtained
 using the tunneling conductivity formula and in \cite{nagaosa2}  it was
found $\rho_{c}\propto \sqrt{T} $.  We will adopt  Kubo formula for
the conductivity as in \cite{nagaosa2}. \\
\indent It
is instructive  to compare  the c-axic conductivity of   Fermi liquid  with
that of  (2D) gauge theories. In Fermi liquid
electrons in a layer are quasiparticles with  some relaxation time and
their retarded Green function has a
pole in the lower half plane. Then, provided that there is no
interlayer scattering, the   Eqn.(\ref{eq;c2}).
 yields
\be
 \sigma_c^{(0)}\sim e^2\, \tc^2 \, m_F\, \tau
\ee
 Therefore, $ \rho_{c} $ and  $\rho_{ab}$ have  the same temperature
dependence.

\indent The situation is  different if the electron is not  quasi particle.
Once the hopping is treated as a perturbation,  electron always decays
to true  quasi-particles during the interlayer tunneling, so the
quantum states of the electron
in different layers are incoherent. As a result of this incoherence, the
out-of-plane transport is blocked and may be relaxed by thermal processes
, which is similar to a semiconducting behaviour.

 The above case  is true of doped Mott insulator: at sufficiently high
temperature electrons decay very fast ( in the time
scale of $1/J$ or $1/\ta$ ) into "spinons" and "holons" and don't
constitute stable excitation. At this temperature range
 the gauge interaction is perturbative and
 electron Green function  is simply a product
of  fermion and boson Green functions ( recall $\,
c_{\alpha}=f_{\alpha}b^{\dag}\, $)
\begin{equation}
\label{GG}
G_e(x,y)=-<f(x)f^{\dag}(y)b(y)b^{\dag}(x)>\sim G_F(x,y)G_B(y,x)
\end{equation}
Therefore, the propagating character of fermion Green function  \\
$ < f(x) f^{\dag}(y) > \sim e^{ \left (i (|\vx-\vy|-v_F(t_x-t_y) \right )}$
is blocked  by  the localized  boson Green function $~<b(y)b^{\dag}(x)>\sim
\frac{T_0}{T}\,\,\exp(-|\vx-\vy|^2 \,m_B T) \, $.

\indent The simplest way to evaluate the integral (\ref{eq;c2})  in gauge
theory is to rewrite it
in the form
of fermi and bose density-density correlation functions $\pi_F$ and $\pi_B$
using the
decomposition Eq.(\ref{GG}).
$$\pi_F(i\omega,{\bf q})=\sum_{\vx} \int _{0}^{\beta} d\tau e^{i{\bf
q}\!\cdot \vx+i\omega \tau}\;G_F^{(t)}(\vx,\tau) G_F^{(b)}(-\vx,-\tau)$$
$$\pi_B(i\omega,{\bf q})= \sum_{\vx}\int _{0}^{\beta} d\tau e^{i{\bf
q}\!\cdot \vx+i\omega \tau} \;G_B^{(t)}(-\vx,-\tau) G_B^{(b)}(\vx,\tau)$$
The superscripts of  Green functions denote two layers involved in
hopping process ( top, bottom). In terms of $\,\pi_F, \;\;\pi_B\,$
(\ref{eq;c2})
takes the form:
\begin{equation}
\sigma_{c}^{(0)}=\,2e^2 \tc^2 \sum_{\bf q} \int
\frac{d\omega}{2\pi}\left(-\frac{\partial\, n_B(\omega) }{\partial
\omega}\right) \,\,\mbox{Im}\,\pi_F^R(\omega,{\bf
q})\,\,\mbox{Im}\,\pi_B^R(\omega,{\bf q})
\label{eq:c1}
\end{equation}
At small frequency and momentum  and at $\,T > T_0 \,$ the imaginary parts of
the polarization
operators are
$$\mbox{Im}\,\pi_F^R(\omega,{\bf q})=-m_F\, a^2 \,\frac{\omega}{v_F
|\vq|}~,~\mbox{Im}\,\pi_B^R(\omega,{\bf q})=-\frac{T_0}{T}\, m_B\, a^2
\frac{\omega}{v_B |\vq|}$$
 , where $a$ is the lattice constant in a layer and $v_B=(k_B T/m_B)^{1/2}~
$ is the thermal boson velocity.
The momentum integration  in Eq.(\ref{eq:c1}) is logarithmic and is cut by
$ T/v_F \;$ at lower limit. Due to  the exponential decay the bose factor
($\, \partial n_B(\omega)  /\partial \omega \propto e^{-|\omega|/T} \;$ ) the
frequency
integral is convergent at ultra-violet limit . The main contribution to the
frequency integral comes from  the region $ |\omega| \le T \;$, in which
$- \partial n_B(\omega)  /\partial \omega\approx T/\omega^2 \;$. The $\omega^2
\,$ in
denominator is cancelled by $\omega^2 \,$ coming from $\,\mbox{Im}\,\pi_F^R
\; \mbox{Im}\,\pi_B^R \,$. Thus the frequency integral gives $\, T^2 \,$.
Rearranging other factors, within logarithmic accuracy, we obtain
\begin{equation}
\label{eq:sig0}
\sigma_{c}^{(0)}=\mbox{const.}\,e^2\, \tc^2 ~ x~ m_F^2 \sqrt{m_B T}
\end{equation}

  The dimensional  crossover to the anisotropic three dimensional regime is
complex.
In particular it evolves hopping process  (\ref{H2D+}) into anisotropic
gauge theory (\ref{H3D}) and
requires
 more sophisticated analysis. Let us
just note that two models (\ref{H2D},\ref{H2D+} )  and (\ref{H3D}) are
essentially different,
so  that an estimate of the crossover temperature from the high temperature
side may not necessarily coincide with the estimate  from the low
 temperature side.
In any case it is very likely that the cross-over temperature in
optimally doped cuprates falls below the superconducting transition
temperature.

A comment is necessary at this point.
In optimal \lan, near $\,T=\mbox {300 K}\,$   the out-of-plane resistivity
stops decreasing and starts to grow with temperature.  This up-turn
is attributed to the structural transformation from high temperature tetragonal
phase to low tempearature orthorhombic phase \cite{kwo2}. Above this up-turn
temperature the c-axis-conductivity is still much lower
than  the Mott minimal metallic conductivity($\approx 10^2 \,\mbox{s/cm}$)
and can not be considered to be metallic.

\section{inelastic scattering by gauge fields}
Two dimensional gauge theory gives rise to  the linear temperature
dependence of  in-plane resistivity of  optimally doped \lan in the regime
where  scattering is {\em elastic}.
At sufficiently high   temperature inelastic processes change the linear-T
behaviour into $ \, T^{3/2} \,$. \\
\indent In three dimensional case the scattering by $\, D_{\perp}(\vk,\omega)
\,$
is  almost always elastic (see Sec. III), while the scattering by $\,
D_{\parallel}(\vk,\omega) \,$ can be inelastic at high temperature. It
turns out that $ \,T^{{\rm 3D}}_{{\rm in}} \, $ is very close to $
\,T^{{\rm 2D}}_{{\rm in}} \, $ .
For the three dimensional inelastic regime to be observed  the condition $
T^{{\rm 3D}}_{{\rm in}}\le T_{{\rm d}}(x)\, $ should be satisfied. ( See
the discussion below on the experimental estimate of $\, T^{{\rm 2D}}_{{\rm
in}} , \, T^{{\rm 3D}}_{{\rm in}} \,$).
In three dimensional inelastic regime we would have $ \,\rho_{ab} \propto
T^{3/2} (\Mb T)^{1/2} \propto T^2 \,$ , so the anomalous
exponent 3/2 cannot be explained. Instead we will discuss two dimensional case
in detail. \\
\indent From the propagator of the gauge field
$~D^{R}(\omega,\vk)=\Bigl(\chi\,k^2-i\,\gamma\,\omega/k\,\, \Bigr)^{-1}~$
it follows that the energy transfer $\omega \,$ scales like $\,
\omega_{\vk}=\chi\gamma^{-1}\,k^3 \,$.
At finite temperature the  boson energy is typically of order $T\,$. Thus
the typical momentum transfer in the scattering  of boson by gauge field is
$ (m_B \, T)^{1/2} \;$. As a result the typical energy transfer in
scattering would be
$ \; \omega*\sim\chi/\gamma\,(m_B T)^{3/2}  \; $. This is larger than
 the thermal  energy of scattered
bosons, i.e.
 $T$ at $\,T >  T_{{\rm in}} \sim \left(\frac{\gamma}{\chi} \right
)^2\frac{1}{m_B^3} \,$
and at this temperature the $\omega\,$ dependence of the propagators has to be
taken into account. This inelasticity softens the
infrared singularity of scattering, thus leads to the less singular
temperature dependence of resistivity.  \\
In 2D case the Eq.(\ref{tau}) reads:
\be
\label{eq:inel}
 \tau^{-1}_{{\rm tr}}(\vp)\sim\pi\int\frac{d^{2}\vk}{(2\pi)^{2}}
\,\frac{|\vp\times\hat{\vk}|^{2}}{m_{B}^{2}}\,\int_{0}^{\infty}\frac{d\omega
}{\pi}\,{\text{Im}}\,D^{R}(\omega,\vk)
(1+n_B(\omega))\,\delta(\xi_{\vp}-\xi_{\vp+\vk}-\omega)\,
\frac{|\vk|^2}{|\vp|^2}
\ee
After angular integration it becomes
\be
\tau^{-1}_{{\rm tr}}\approx \frac{1}{v_{\vp} \, m_B^2 \, \gamma} \int_0^{|\vp|}
k^3 dk
\int d \omega  \,\frac{\omega}{(\omega^*_{\vk})^2+\omega^2}
\;\left(1+n_B(\omega/T) \right)
\ee
 At low temperature  $\, T <
T_{{\rm in}} \,$,  $\, \omega \sim \omega^*_{\vk} \ll T \,$ thus $\,
n_B(\frac{\omega}{T}) \sim T/\omega  \gg 1\,$. Then the frequency integral
is finite and it gives $\, (\omega^*_{\vk})^{-1}\,$. The remaining momentum
integration gives T-linear transport time $\,
\tau^{-1}_{{\rm tr}} \sim T/( \chi \, m_B ) \,$ \cite{lee3,wieg1}.  Note
that the transport time is indepedent of Landau damping parameter $\gamma$
, which is not the case in inelastic regime. \\
\indent At high temperature  $\, T > T_{{\rm in}} \,$ $\, \omega \sim
\omega^*_{\vk} \gg T \,$ thus $\, n_B(\frac{\omega}{T}) \ll 1 \,$. Now the
frequency integral is the order of $\, \log \Lambda \,$ , where $\,\Lambda \,$
is some high frequency cut-off.  The
momentum integral gives $ p^4 \, $.
Combining all factors and repalcing the boson momentum by its thermal value $\,
(m_B T)^{1/2} \,$ we obtain  in inelastic limit,
\be
\tau^{-1}_{{\rm tr}}\approx \frac{p^3}{ m_B \, \gamma}  \propto
\frac{T^{3/2} m_B^{1/2} }{\gamma}
\ee

\indent The value of  $T_{{\rm in}}\,$  is very sensitive to $\,m_B\,$ and can
hardly be estimated from the available experimental data.  The slope of
T-linear
 resistivity  at optimal doping ( $ \,\approx 1.0  \;\mu \,\Omega \,{\rm
cm}/{\rm K} \,$) gives $\chi^{{\rm 2D}}\,$ to be around 500 K.
The resistivity data of  overdoped \lan  suggest that $\chipa^{{\rm 3D}}\,$  is
the same order as  $\chi^{{\rm 2D}}\,$.  The one-loop value of the damping
$\gamma$ is order of 1 \cite{kotliar}.  \\
\indent The estimates of $m_B$ which enters into $T_{{\rm in}}\,$ vary
appreciably depending on the
kinds of experiments.  The
optical conductivity measurements \cite{uchida3} providess the value of $m_B$
at high energy : $m_B\approx 2 \, m_e \,$, which is almost independent of
doping. Especially $m_B\approx (2 \sim 3 ) m_e \,$ is almost independent
of the probe energy scale {\em in overdoped range}.
 From another side the magnetic susceptibility data provides the value of
$m_B$
at low energy :   $m_B\approx 15\,  m_e
\;$ near optimal doping\cite{walsted,kalmeyer}.
  These estimates of $m_B$ makes  the estimate of $T_{{\rm in}}\,$ range from
$500 \,\mbox{K}\,$ ( the susceptibility data) to $10^5 \,\mbox{K}\,$ ( the
optical data). \\
\indent If $\,T_{{\rm d}}(x) \le T_{{\rm in}} \, $ for some doping range the
following behavior of the resistivities are possible.
\ba
\rho_{ab}\sim \rho_c &\propto & T^{3/2} ,   \quad {\rm for}
\quad T \le T_{{\rm d}}  \le  T_{{\rm in}} \nonumber \\
\rho_{ab} &\propto & T , \;\; \rho_c \propto T^{-1/2}  \quad {\rm for} \quad
T_{{\rm d}} \le T \le T_{{\rm in}} \nonumber \\
\rho_{ab} &\propto & T^{3/2} ,   \quad {\rm for}
\quad  T_{{\rm d}} \le T_{{\rm in}} \le T
\ea
If one accepts the lower estimate of $\,T_{{\rm in}}\,$ one may exploit the
inelastic mechanism in order to explain $T^{3/2}\,$ behaviour.
 If  $\, T_{{\rm d}}(x) \ge T_{{\rm in}} \, $
\ba
\rho_{ab} \sim \rho_c &\propto & T^{3/2} ,   \quad {\rm for}
\quad T \le T_{{\rm in}}  \le T_{{\rm d}} \nonumber \\
\rho_{ab} \sim \rho_c &\propto & T^2 ,   \quad {\rm for} \quad
T_{{\rm in}} \le T \le T_{{\rm d}} \nonumber \\
\rho_{ab} &\propto & T^{3/2} ,   \quad {\rm
for} \quad T_{{\rm in}} \le  T_{{\rm d}} \le T
\ea
\indent In fact the optical estimate , which is close to the band theory value,
seems
more realistic. This means that the inelstic regime is very likely irrelevant
for the overdoped cuprates.

\section{Conclusion}
We  adopted the gauge theory of normal states of doped Mott insulator
to explain anomalous transport phenomena observed in overdoped cuprates. We
assumed
that  \lan interpolates between layered and  anisotropic anomalous metal
for the  doping range $x \sim 0.15-0.35$ and still does not evolve into the
ordinary metallic behaviour. We attempted to describe the  dimensional
crossover of
the anomalous metal in temperature. The crossover of the out-of-plane transport
is peculiar: due to  strong interaction electrons do not constitute
an elementary excitation and decay into other particles during the interlayer
tunneling. As a result, the  character of the out-of-plane transport may
change from coherent to incoherent and that of the out-of-plane resistivity
changes from metallic to semiconductor-like behaviour.  In addition we
discussed another
crossover between elastic and inelastic scattering as  temperature
increases. The theory provides an unified approach in understanding the variety
of temperature behaviours of the in-plane and out-of-plane resistivity  of
cooper oxides in wide ranges of doping and temperature. The results
qualitatively
agree with the available experimental data.


\vskip 0.5cm
\centerline{\bf ACKNOWLEDGEMENTS}
We would like to thank  K. Levin for stimulating our interest
to the problem of incoherent transport in highly anisotropic systems and
very useful discussions. We acknowledge that L. B. Ioffe obtained results
of this paper independently and thank him for the collaboration in the
first stage of the project.  \\
P. W. thanks B. Battlog and L. Cooper for useful discussions.
H. L. is also grateful to A. Abanov, Y. B. Kim and P. Solis for many
discussions.
This paper was first presented in APS  meeting in 1994. \\
Authors were supported in part by the National Science Foundation (DMR
91-20000)
through the Science and Technology Center for Superconductivity.



\begin{references}

\bibitem{kwo1} H. Takagi, {\it et al.}, Phys. \ Rev. \ Lett. \ {\bf 69},
2975 (1992).
\bibitem{kwo2} H. L. Kao, J. Kwo, H. Takagi, and B. Batlogg,
Phys. \ Rev. \ B \ {\bf 48} , 9925 (1993)

\bibitem{uchida1} Y. Nakamura and S. Uchida,
Phys. \ Rev. \ B \ {\bf 47}, 8369 (1993)
\bibitem{uchida2} K. Tamasaku, T. Ito, and S. Uchida,
Phys. \ Rev. \ Lett. \ {\bf 72}, 3088 (1994)
\bibitem{uchida3} S. Uchida, {\it et al.},
Phys. \ Rev. \ B \ {\bf 43}, 7942 (1994)
\bibitem{walsted} R. E. Walsted, {\it et al.},
Phys. \ Rev. \ B \ {\bf 45}, 8074 (1992)
\bibitem{iye} Y.  Iye,
in {\it Physical Properties of High Temperature Superconductors III}, \\
 edited by D. M. Ginsberg (World Scientific, 1992) .
\bibitem{cooper} S. L. Cooper and K. E. Gray,  in {\it Physical Properties
of High Temperature Superconductors IV}, edited by D. M. Ginsberg (World
Scientific, 1995)
\bibitem{kwak} J. F. Kwak,
Phys. \ Rev. \ B \ {\bf 26}, 4789 (1982)
\bibitem{allen} P. B. Allen, W. E. Pickett, and H. Krakhauer,
Phys. \ Rev. \ B \ {\bf 36}, 3926 (1987)
\bibitem{pickett} W. E. Pickett, Rev. Mod. Phys. {\bf 61}, 433 (1989)
\bibitem{note1} Models where an interlayer scattering (which does change the
character
of  quasiparticles) is larger than the interlayer hopping amplitude and
in-plane scattering rate  are considered in Refs.\cite{graf,levin,shapiro}.
In these papers  the behaviour $\rho_{ab} \propto 1/\rho_c~$ was found, so that
the out-of-plane resistivity $\rho_c$ decreases with increasing temperature.


\bibitem{lee1} P. A. Lee,
Phys. \ Rev. \ Lett. \ {\bf 63}, 680 (1989)
\bibitem{lee3} N. Nagaosa and P. A. Lee,
Phys. \ Rev. \ Lett. \ {\bf 64}, 2450 (1990) ;
 P. A. Lee and N. Nagaosa,
Phys. \ Rev. \ B \ {\bf 46}, 5621 (1992)
\bibitem{nagaosa2} N. Nagaosa,
J.\ Chem.\ Phys.\ Solids \ {\bf 53}, 1493 (1992)
\bibitem{larkin} L. B. Ioffe and A. Larkin,
Phys. \ Rev. \ B \ {\bf 39}, 8988 (1989)
\bibitem{kotliar} L. B. Ioffe and G. Kotliar,
Phys. \ Rev. \ B \ {\bf 42}, 10348 (1990)
\bibitem{wieg1} L. B. Ioffe and P. B. Wiegmann,
Phys. \ Rev. \ Lett. \ {\bf 65}, 1653 (1990)
\bibitem{kalmeyer} L. B. Ioffe and V. Kalmeyer,
Phys. \ Rev. \ B \ {\bf 44}, 750 (1992) ;
M. V. Feigelman, V. B. Geshkenbein, L. B. Ioffe, and A. I. Larkin,
Phys. \ Rev. \ B \ {\bf 48}, 16641 (1993)
\bibitem{anderson} P. W. Anderson and Z. Zou,
Phys. \ Rev. \ Lett. \ {\bf 60}, 132 (1988) ; P. W. Anderson, {\it Princeton
RVB Book~}, (unpublished) Chapter VI.

\bibitem{anderson2} P. W. Anderson, Science {\bf 256}, 1526 (1992)
\bibitem{anderson3} S. Chakravarty, A. Sudbo, P. W. Anderson, and S.
Strong, Science {\bf 261}, 337 (1993)
\bibitem{kumar} N. Kumar and A. M. Jayannavar,
Phys. \ Rev. \ B \ {\bf 45}, 5001 (1992)
\bibitem{legget}  A. J. Leggett,
Braz. \ J. \ Phys. \ {\bf 22}, 129 (1992)
\bibitem{graf} M. J. Graf, D. Rainer, and J. A. Sauls,
Phys. \ Rev. \ B \ {\bf 47}, 12089 (1993)
\bibitem{levin} A. J. Rojo and K. Levin,
Phys. \ Rev. \ B \ {\bf 48}, 16861 (1993)
\bibitem{shapiro} N. Kumar, P. A. Lee, and B. Shapiro,
Physica A {\bf 168}, 447 (1990)
\end{references}
\end{document}